# Appearance and disappearance of superconductivity with Fe site Co substitution in $SmFe_{1-x}Co_xAsO$ (x = 0.0 to 1.0)

V.P.S. Awana, Anand Pal, Mushahid Hussain and H. Kishan

*National Physical Laboratory(CSIR), Dr. K.S. Krishnan Marg, New Delhi-110012, India*
*Deprtment of Physics, Jamia Millia Islamia, Central University, New Delhi-110025, India*

**Abstract**

We report synthesis, structural details and magnetization of $SmFe_{1-x}Co_xAsO$ with x ranging from 0.0 to 1.0 at close interval of 0.10. It is found that Co substitutes fully at Fe site in SmFeAsO in an iso-structural lattice with slightly compressed cell. The parent compound exhibited known spin density wave (SDW) character below 150K. Successive doping of Co at Fe site suppressed the SDW transition for x = 0.05 and later induced superconductivity for x = 0.10, 0.15 and 0.20 respectively at 14, 15.5 and 9K. The appearance of bulk superconductivity is established by wide open isothermal magnetization M(H) loops. For higher content of Co i.e. x ≥ 0.30, superconductivity is not observed. Clearly the Co substitution at Fe site in $SmFe_{1-x}Co_xAsO$ diminishes the Fe SDW character (x=0.05), introduces bulk superconductivity for x from 0.10 to 0.20 and finally becomes a non-superconductor. The SmCoAsO also exhibits a secondary AFM like transition at below around 50 K. The reported AFM ordering of Sm spins is seen from heat capacity Cp(T) at 4.5 K and the same remains invariant with Co doping in $SmFe_{1-x}Co_xAsO$. Further the FM ordering of Co-spins (seen in magnetization measurements) is not evident in Cp(T) studies, suggesting weak correlations between ordered Co-spins in these systems.

[*] Corresponding Author

Dr. V.P.S. Awana
National Physical Laboratory, Dr. K.S. Krishnan Marg, New Delhi-110012, India
Fax No. 0091-11-45609310: Phone No. 0091-11-45608329
e-mail-awana@mail.nplindia.ernet.in: www.freewebs.com/vpsawana/

## 1. Introduction

The recent discovery of superconductivity (SC) in F doped LaFeAsO compound reported by Kamihara et al.[1] has been of tremendous interest to the condensed-matter physics researchers. The discovery is fundamentally striking due to the presence of the familiar magnetically ordered atom *iron* being showing superconductivity comparable to high Tc compounds at 26 K. The superconducting transition of this compound increased up to 43 K under pressure of ~4 GPa.[2]. Soon after, the Tc increased up to ~56 K by replacing La ion with other rare earth elements such as Ce, Pr, Sm, Nd etc., which is the highest expect high-Tc cuprates.[3-9].

The parent compound REFeAsO (RE= rare earth) itself is non-superconducting and the same exhibits spin density wave like short range antiferromagnetic (AFM) order clubbed with a structural distortion at around 140K. After doping of carriers, the magnetic order is suppressed and the compound becomes superconducting along with a good metallic behaviour down to $T_C$. The optimal doping for Superconductivity in REFeAsO, was studied by: (i) F substitution on the O sites, (ii) by inducing oxygen deficiency, and (iii) by partial substitution of the trivalent Rare-earth (RE) ions by bi-or tetravalent cationic species [3-9]. Recently it has been shown that superconductivity can be induced in REFeAsO compound by Fe site Co substitution [10-13], which results in direct injection of electrons in the conducting FeAs layers.

Here we report synthesis, structural details, magnetic and electrical properties of $SmFe_{1-x}Co_xAsO$ (x=0.0 to 1.0). It is found that Co substitutes fully at Fe site in SmFeAsO in an iso-structural P4/nmm space group with slightly compressed cell. The parent compound SmFeAsO showed signs of known SDW magnetic order in resistivity versus temperature measurement at around~140 K. Successive doping of Co at Fe site suppressed the SDW transition for x = 0.05 and later induced superconductivity for x = 0.10, 0.15 and 0.20 respectively at 14, 15.5 and 9K. Superconductivity is not seen for higher Co content ( x ≥ 0.30), which is most probably due to over doping of carriers. Fully Co substituted i.e., SmCoAsO exhibited an anomalous magnetic behaviour in comparison to reported LaCoAsO, indicating the interaction between magnetic RE spins and adjacent layer ordered Co spins.

## 2. Experimental Details

Polycrystalline SmFe$_{1-x}$Co$_x$AsO ( x=0.0 to 1.0) samples were synthesized by single step solid-state reaction method via vacuum encapsulation technique. Stoichiometric amounts of high purity (~99.9%) Sm, Fe, As, Co$_3$O$_4$, Fe$_2$O$_3$ and Co powder ground thoroughly using mortar and pestle. All the weighing and grinding is done in the glove box under the controlled atmosphere. The oxygen and H$_2$O content in the glove box is around 1ppm level. The mixed powders were palletized and vacuum-sealed (10$^{-4}$ Torr) in a quartz tube. These sealed quartz ampoule was placed in box furnace and heat treated at 550$^o$C for 12 hours, 850$^o$C for 12 hours and then finally 1150$^o$C for 33 hours in continuum with slow heating rate. Then furnace was allowed to cool naturally. The entire heating process is "Single Step" in nature as reported earlier by some of us [11,14].

The X-ray diffraction pattern of the compound was taken on Rigaku X-ray diffractometer with Cu K$_\alpha$ radiation. The resistivity measurements were carried out by conventional four-probe method on a quantum design *PPMS* (Physical Property Measurement System). Heat capacity and magnetization measurements were also carried out on the same *PPMS*.

## 3. Results and Discussion

The room temperature Rietveld refined fitted X-ray diffraction (*XRD*) pattern of SmFe$_{1-x}$Co$_x$AsO (x = 0.0 to 1.0) compound samples are shown in Fig.1 (a). The simulated Rietveld analyses confirmed that all the studied samples are crystallized in P4/nmm space group with tetragonal structure in analogy of other REFeAsO systems. All the main peaks are indexed with space group P4/nmm and no extra reflections are observed. In tetragonal crystal lattice Sm & As both atoms are located at Wyckoff positions 2c, O is situated at 2a and Fe/Co are shared at 2b site. The observed crystal structure of the Sm(Fe/Co)AsO are shown in Fig.1(b). Sm(Fe/Co)AsO is a layered structure, in which the Fe/Co-As layer is sandwiched between the Sm-O layer. Fig. 1(c) depicts the lattice parameters variation of SmFe$_{1-x}$Co$_x$AsO samples with x ranging from 0.0 to 1.0. It is observed that with Co doping c parameter decreases while a lattice parameter remain nearly unaltered. The decrease in c lattice parameter show evidence of successful chemical substitution of $Co^{3+}$ on $Fe^{2+}$ Site. The incorporation of Co in the Fe site reduces the cell volume due to contraction of the c-axis lattice constant. It is evident that Co substitutes successfully with full solubility at Fe site in SmFeAsO. The details of co-ordinate positions and lattice parameters along with the quality of fit parameters are given in Table 1 and Table 2.

Figure 2(a) shows the temperature versus resistivity behaviour for $SmFe_{1-x}Co_xAsO$ (x=0.0, 0.15, 0.30 and 1.0) and the variation of resistivity under magnetic field ρ(T)H for the same of up to 9 Tesla is shown in the inset of same figure i.e., Fig.2(a). The parent compound SmFeAsO does not show superconductivity and rather shows the metallic step anomaly at near 140 K. The anomaly is due to the spin density wave (SDW) instability and the structural phase transition from tetragonal to orthorhombic phase, as being reported earlier [3,11]. Successive doping of Co at Fe site suppressed the SDW transition for x = 0.05 and later induced superconductivity for x = 0.10, 0.15 and 0.20 respectively at 14, 15.5 and 9K [15]. At room temperature the resistivity value for $SmFe_{0.85}Co_{0.15}AsO$ sample is 3.8 mΩ, which is much smaller as compared to the pure SmFeAsO. The resistivity behaviour of $SmFe_{0.85}Co_{0.15}AsO$ is metallic from room temperature down to superconductivity onset and it shows superconducting transition $T_c$ (R=0) at 14 K. With further doping of Co ≥ 0.30 superconductivity disappears, which is primarily due to the over doping of carriers. Further, it is observed that as the cobalt content increase (≥ 0.30), the metallic behaviour of the compound become more prominent. The increasing metallic behaviour of $SmFe_{1-x}Co_xAsO$ compound evidences the increment of the carrier concentration with Co doping. For example fully Co doped SmCoAsO is metallic altogether down to low temperatures with increased metallic slope of resistivity with temperature than any other studied sample. Although, the superconducting transition temperature for $SmFe_{0.85}Co_{0.15}AsO$ sample from resistivity measurements as ρ=0 is at 14K, the sample showed the diamagnetism onset at slightly higher temperature at 15.5 K, see Fig 4 (to be discussed later). In fact the $T_c^{dia}$ generally coincides with the superconducting transition onset. The R(T) behaviour of SmCoAsO is metallic down to low temperature and it neither exhibit SDW character like SmFeAsO or superconductivity as for $SmFe_{1-x}Co_xAsO$.

The R(T)H measurements for maximum $T_c$ $SmFe_{0.85}Co_{0.15}AsO$ sample are depicted in inset of Fig. 2(a). The R(T)H results reveal that the superconducting transition temperature shifts to lower temperature with applying magnetic field. The transition width for the sample becomes wider with increasing H. From the R(T)H it's clear that rate of decrease of transition temperature with applied magnetic field is 1 Kelvin per Tesla {$dT_c/dH \sim 1K/T$} for Co substituted $SmFe_{0.85}Co_{0.15}AsO$ sample. This value is far less than that of YBCO {$dT_c/dH \sim 4K/T$} and $MgB_2$ {$dT_c/dH \sim 2K/T$} samples; it suggests toward a high value of upper critical field ($H_{c2}$) in these compounds [16]. The upper critical field $H_{c2}$, can be roughly defined as the field at which the superconductor become a normal conductor. Fig.2(b) shows the temperature dependences of upper critical fields $H_{min}$, $H_{mid}$, and $H_{max}$ (all in T) calculated at 10%, 50% and 90% of the normal state resistance at the transition temperature. The fitting of experimental data is done according to the Ginzburg- Landau (GL) theory, which not only determines the $H_{c2}$ value at zero Kelvin [$H_{c2}(0)$], but also determines the temperature dependence of critical field for the whole temperature range. To determine $H_{c2}(0)$ value, the GL equation is:-

$$H_{c2}(T)=H_{c2}(0)*[(1-t^2)/(1+t^2)]$$

Where, $t = T/T_c$ is the reduced temperature and $H_{c2}(0)$ is the upper critical field at temperature Zero. $H_{min}$(10%), $H_{mid}$(50%) and $H_{max}$(90%) are estimated to be 13.5, 21 and 36Tesla respectively for $SmFe_{0.85}Co_{0.15}AsO$ at 0 K.

DC magnetic susceptibility versus temperature plots in both zero-field-cooled (zfc) and field-cooled (fc) of superconducting samples x=0.10, 0.15 and 0.20 are shown in the Fig.3 (a). Inset of fig.3 (a) shows *M(H)* for the $SmFe_{0.85}Co_{0.15}AsO$ sample at 2 K, 5 K and 9 K. The opening of *M(H)* loop give the clear evidence of superconductivity. It's clear from DC susceptibility that x= 0.15 samples show the highest transition temperature. For x = 0.10 the sample is under doped, on the other hand for x =0.20 the same is over doped. For samples with x > 0.25 superconductivity is not observed due to over doping. $Fe^{2+}$ site $Co^{3+}$ substitution provides effective e type carriers to the doped samples.

The ZFC and FC magnetization of fully Co substituted i.e., SmCoAsO sample in an applied field of 10 Oe is shown in Fig. 4(a). The isothermal magnetization M(H) plots of the compound at various temperatures of 150, 75, 50, 30 and 15K are shown in inset of Fig. 4(a). The SmCoAsO is paramagnetic 75 K and exhibits ferromagneic (FM) behaviour below this temperature. This is clear from the fact that the M(H) are linear (inset Fig. 4(a)) below 75K. At 50K compound exhibits good ferromagnetic behaviour via saturated M(H) loops as seen in inset of Fig. 4(a). Seemingly the Co spins ferromagnetically below say 75K. This is similar to that as reported earlier for LaCoAsO, where Co spins order FM below about 100 K [10]. Interestingly below 50 K, the compound exhibits a down turn in both FC and FC magnetic susceptibility, indicating towards anti ferromagnetic (AFM) correlations. In fact there are two peaks below in both ZFC and FC just above and below 50K indicating towards competing FM and AFM. Further the M(H) plots at 30K show some component of meta magnetic transition in an spin glass (SG) like S-type loop with FM like saturation above 1 Tesla field, see inset Fig. 4a. This is very different that as reported for non magnetic RE ion (La) containing LaCoAsO, where the Co spins order FM below say 100K and remain FM down to low temperatures [10]. It is clear that in SmCoAsO, the Co spins order FM below 75 K, undergoes a SG transition just around 50 K, develops a meta magnetic state and finally order AFM below say 30 K.

To further understand the magnetic behaviour of SmCoAsO, we carried out the FC magnetization of SmCoAsO at higher fields (1 Tesla), which is depicted in Fig. 4(b). The 1/M versus temperature plot of the same is given in inset of Fig. 4(b). At higher fields (1 Tesla) the compound goes through a PM-FM-AFM transition without the SG component. This because the two peaks (Fig. 4a) seen around 50K in ZFC/FC at 10 Oe are not evident. The 1/M versus T plot shown in inset of Fig. 4(b), clearly shows that SmCoAsO is PM above 100K, FM between 30K to 80K and finally AFM below 30K. This gives credence to the fact that magnetic ion Sm influences the ordering of Co ions in adjacent CoAs layer. This different to that as reported widely for case of $REBa_2Cu_3O_7$ (RE=rare earth), where the superconductivity appears at 90K, irrespective of the nature (non magnetic/magnetic) of RE, except for the case of Pr. In case of Pr the $Pr^{4f}$ orbitals do hybridise with the superconducting plane $O^{2p}$ orbital. Interestingly in case of Fe based doped SmFeAsO/F the superconductivity is seen and that also at much higher temperature than as for doped LaFeAsO/F. Thus indicating that magnetic Sm ions though effect the ordering of Co ions in adjacent CoAs layer, the same does not effect the superconductivity in same structured FeAs layers. Interestingly, the intriguing magnetization of SmCoAsO is quite similar in qualitative way to another magnetic RE containing similar compound NdCoAsO [17]. It is note worthy here that magnetizations measurements do not necessarily land the true picture of the magnetic structure of the compound, but gives an idea about possible magnetic ordering. What is clear from our magnetization results being shown in Fig. 4(a, b) is that the ordering of Co in SmCoAsO is very different than as reported for LaCoAsO [10] and qualitatively close to that as for NdCoAsO [17].

The zero field Heat capacity ($C_p$) versus temperature for the both the end sample of the series, SmFeAsO and SmCoAsO are shown in main panel of fig. 5. At 200 K Heat capacity ($C_p$) reaches the value around 3.5 R. which is comparable to the previous report [17]. Both the curve overlap to each other in the entire temperature range, except for the distinct kink around 140 K observed for SmFeAsO. The small kink in SmFeAsO heat capacity versus temperature plot around 140 K is due to the Spin density Wave (SDW) character of the compound. Resistivity versus temperature measurements also shows the metallic step at 140 K. Another peak is observed at 4.5 K in both the samples, which is due to the antiferromagnetic orderings of $Sm^{3+}$ ions. The peaks in specific heat shown in Fig. 5 were integrated to get entropy change associated with each transition. To calculate the non-magnetic contribution of heat capacity for each transition, we are using the polynomial interpolation fitting above and below the transition [17]. This is because of the absence of an appropriate non-magnetic analogue for the background lattice/electronic contribution to the heat capacity. To estimate the background contribution in the specific heat for the SDW transition in SmFeAsO sample, we are using the equation $aT+bT^3$ for the fitting purpose. We fit this equation in the temperature range 160 K to 100 K. Using the fitted value of coefficient $a$ and $b$, we calculate the background curve in the whole temperature (160 K to 100 K) range. The fitted and the experimental curve for the SDW transition in SmFeAsO compound around 135 K are shown in inset-I of Fig 5. To calculate the change in specific heat ($\Delta C_P$) subtracted background contribution from the experimental data and $\Delta C_P/ T$ was integrated to obtain the entropy change $\Delta S$. The integrated result in $\Delta S = \sim 1.34 \times 10^{-2}$ R for the SDW transition in SmFeAsO sample. Same interpolation scheme were applied for the other transition. The heat capacity transition for SmFeAsO at around 4.5 K is shown in inset-III in the Fig.5. To elucidate the entropy change for $C_P$ transition at 4.5K, we fit the above equation in temperature range from 20 K to 2 K. Appling the same process we estimate the entropy change $\Delta S = \sim 0.1782$ R , which is comparable to the same transition in the NdFeAsO [17]. With the same proposal we calculate the entropy change $\Delta S =\sim 0.1412$ R in SmCoAsO sample around 5 K transition. It is clear that $\Delta S$ associated with Sm spin's AFM ordering at 4.5K is clearly more for SmFeAsO than SmCoAsO. This again indicates that Sm magnetic interactions with adjacent CoAs are different than as for FeAs in same structure SmCoAsO and SmFeAsO respectively.

Summarily, our results highlight the fact that Co substitution at Fe site in SmFe$_{1-x}$Co$_x$AsO diminishes the Fe SDW character (x=0.05), introduces bulk superconductivity for x from 0.10 to 0.20 and finally becomes a non-superconductor. Further the magnetic behavior of the end member SmCoAsO is very different than that of LaCoAsO, indicating possible influence of RE ion (Sm) moment on ordering Co in adjacent CoAs layer.

**Figure caption:-**

Figure 1(a): Rietveld fitted room temperature X-ray diffraction patterns of SmFe$_{1-x}$Co$_x$AsO; x= 0.15, 0.20,0.50 & 1.0 compounds.

Figure 1(b): Variation of lattice parameter and unit cell volume with of Co concentration.

Figure 1(c): Arrangement of the atom in Sm(Fe/Co)AsO Unit cell

Figure 2(a): Resistivity behavior with temperature variation $\rho(T)$ of SmFe$_{1-x}$Co$_x$AsO; x=0.0,0.15 & 1.0 sample at zero field. Inset shows the variation of resistivity in the presence of applied magnetic field $\rho(T)H$ up to 9 Tesla.

Figure 2(b): Ginzburg- Landau (GL) equation fitted Upper critical field for the sample SmFe$_{0.85}$Co$_{0.15}$AsO at 90%,50%and 10 % drop of resistance of the normal state resistance

Figure 3(a): Temperature variation of magnetic susceptibility M(T) in FC and ZFC condition for SmFe$_{1-x}$Co$_x$AsO; x=0.10, 0.15 & 0.20 compounds. Inset shows complete magnetization loops M(H) at 2 K for the same.

Figure 3(b): Temperature dependence of magnetic susceptibility M(T) in FC and ZFC condition for SmFe$_{0.85}$Co$_{0.15}$AsO sample. Inset (i) shows the complete magnetization loops M(H) at 2, 5 & 9 K for the same and inset (ii) shows the first quadrant of M(H) loop at 2, 5, 7 & 9 K for SmFe$_{0.85}$Co$_{0.15}$AsO sample.

Figure 4(a) DC magnetic moment versus temperature plots for SmCoAsO in both FC and ZFC situations at 10 Oe. The inset shows the isothermal magnetization M(H) plots for the same at varying temperatures.

Figure 4(b) DC magnetic moment versus temperature plot for SmCoAsO in FC situation at 1Tesla field, the inset shows inverse of the same.

Figure 5 Zero field heat capacity versus temperature C$_P$(T) plots for both SmFeAsO and SmCoAsO, the inset I, II and II shows respectively the C$_P$(R) T for SmFeAsO near SDW temperature and same for SmCoAsO and SmFeASO near Sm ordering temperatures.

Table 1(a); Wyckoff position for Sm(Fe/Co)AsO

| Atom | Site | X | Y | Z |
|---|---|---|---|---|
| Sm | 2c | 0.25 | 0.25 | 0.135(6) |
| Fe/Co | 2b | 0.75 | 0.25 | 0.00 |
| As | 2c | 0.25 | 0.25 | 0.651(4) |
| O | 2a | 0.75 | 0.25 | 0.00 |

Table 1(b); Reitveld refine parameters for SmFe$_{1-x}$Co$_x$AsO (x =0.0 – 1.0) with space group P4/nmm.

| x | $a\ (A)$ | $b\ (A)$ | $c\ (A)$ | Vol.(A$^3$) | R$_{wp}$ | chi$^2$ |
|---|---|---|---|---|---|---|
| 0 | 3.9372(3) | 3.9372(3) | 8.4980(11) | 131.73(2) | 4.34 | 3.09 |
| 5 | 3.9403(3) | 3.9403(3) | 8.4823(10) | 131.70(2) | 4.38 | 2.34 |
| 10 | 3.9402(4) | 3.9402(2) | 8.4647(13) | 131.41(3) | 5.40 | 3.18 |
| 15 | 3.9391(2) | 3.9391(2) | 8.4672(6) | 131.38(1) | 4.36 | 2.80 |
| 20 | 3.9404(3) | 3.9404(3) | 8.4449(9) | 131.12(2) | 4.65 | 2.54 |
| 30 | 3.9393(2) | 3.9393(2) | 8.4217(6) | 130.69(1) | 4.15 | 2.10 |
| 50 | 3.9382(1) | 3.9382(1) | 8.3872(5) | 130.08(1) | 3.79 | 2.29 |
| 80 | 3.9461(4) | 3.9461(4) | 8.3227(12) | 129.60(2) | 4.55 | 2.72 |
| 100 | 3.9573(4) | 3.9573(4) | 8.2411(12) | 129.06(2) | 4.93 | 2.34 |

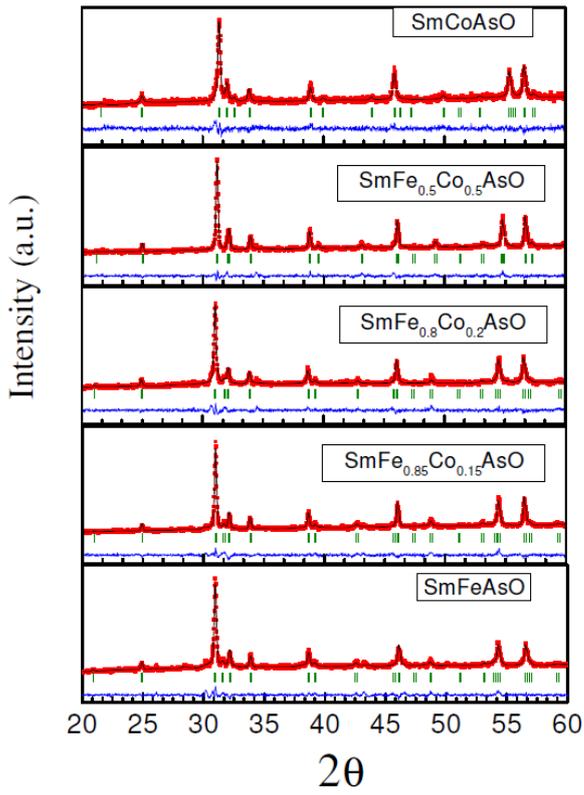

Fig.1(a)

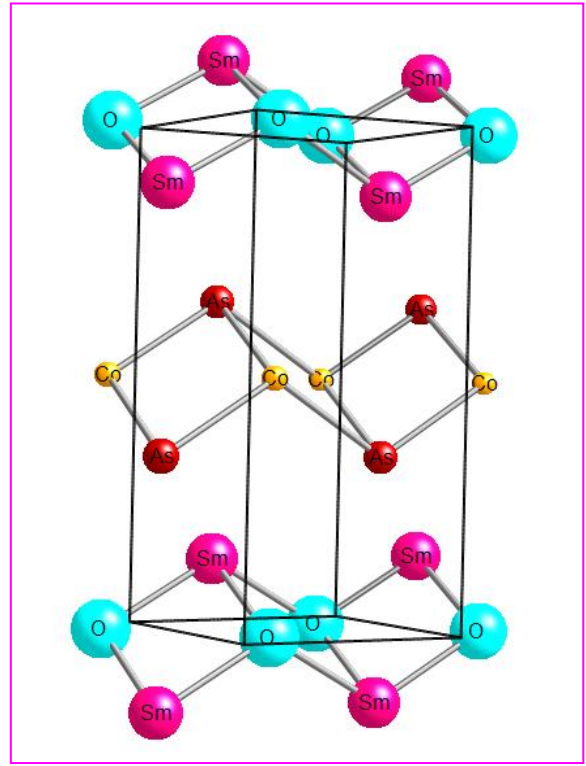

Fig. 1(b)

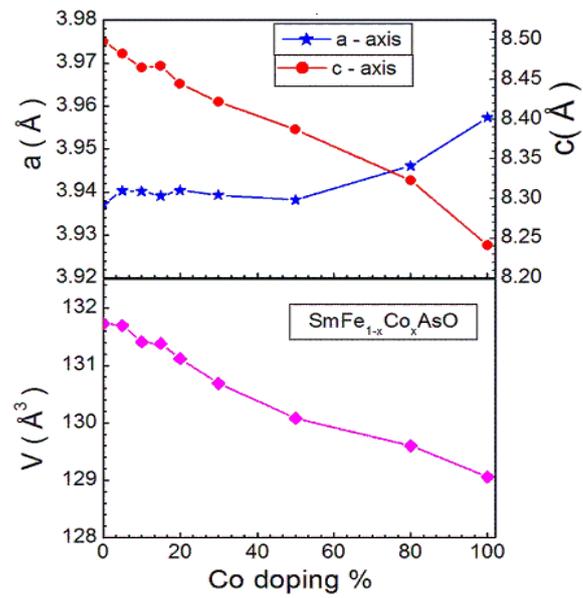

Fig.1(c)

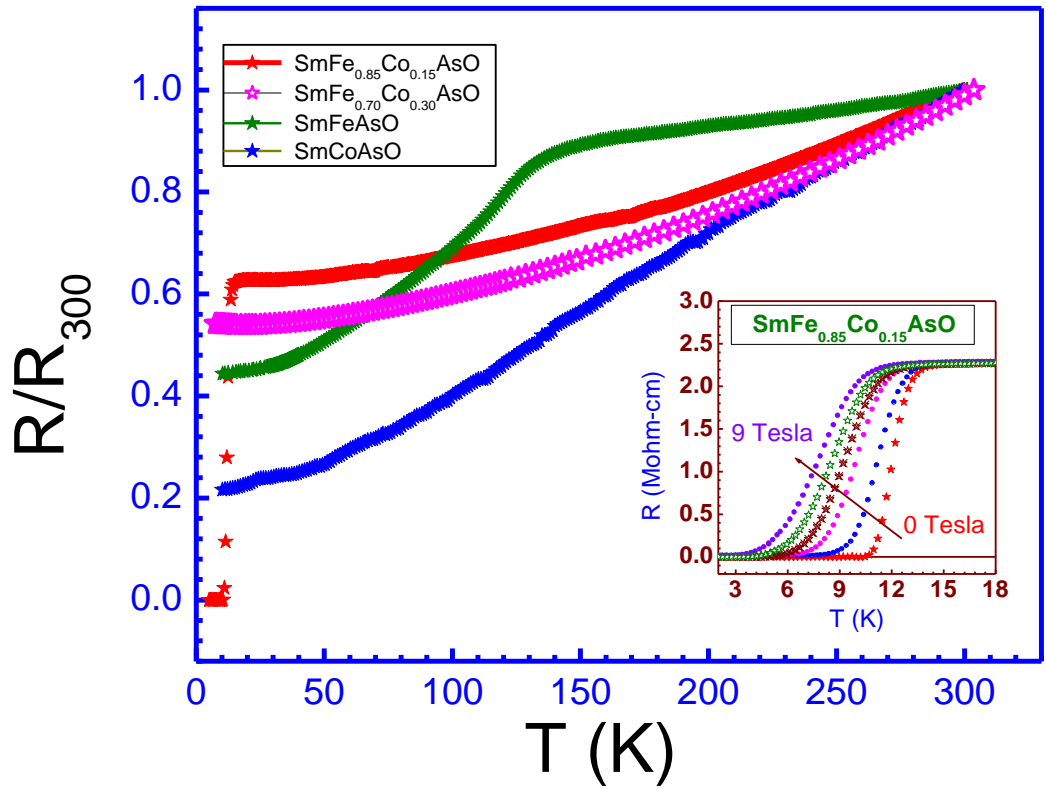

Fig.2 (a)

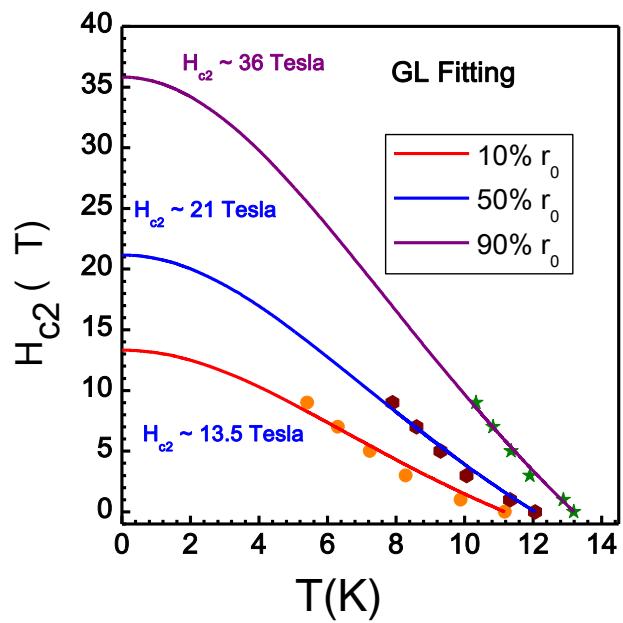

Fig.2(b)

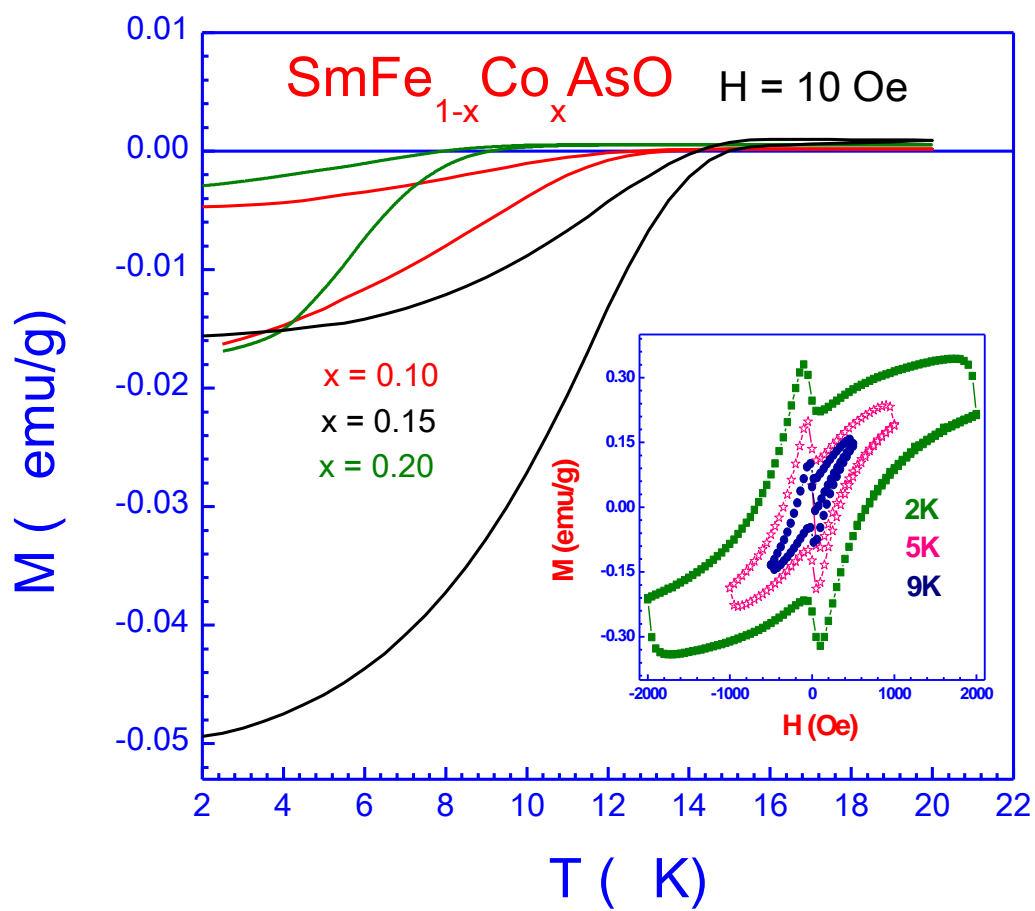

Fig.3

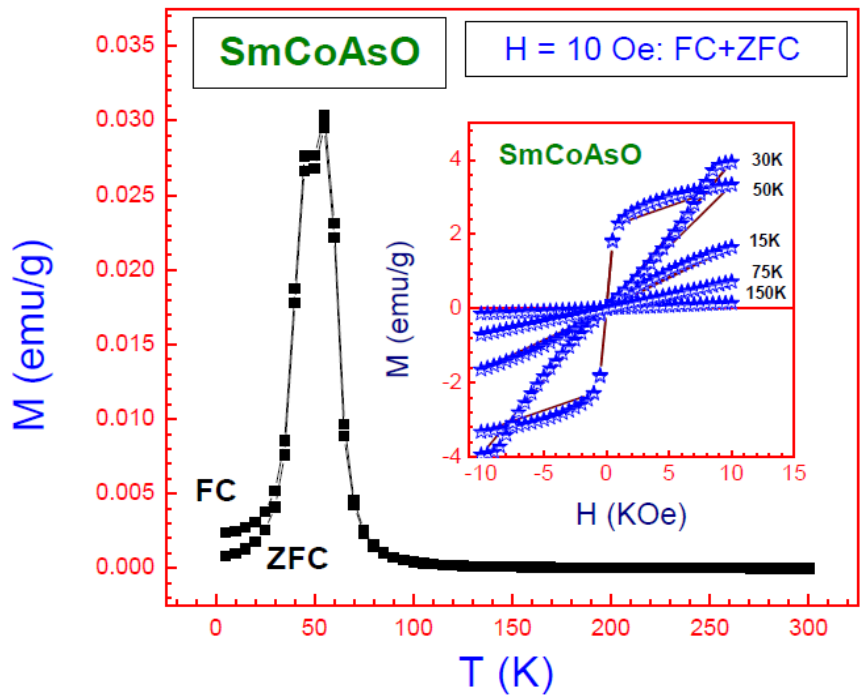

Fig. 4(a)

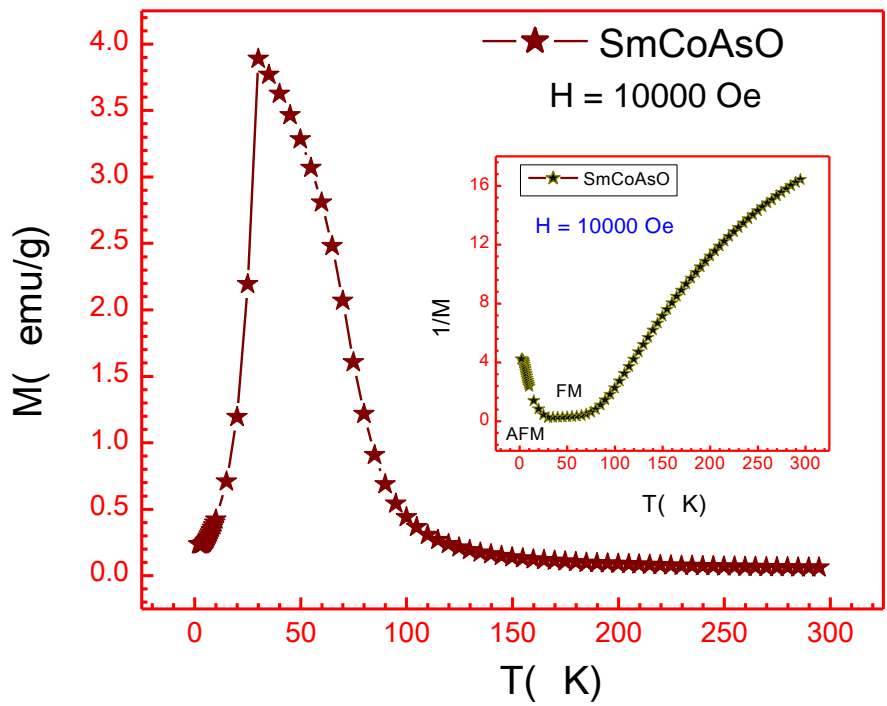

Fig. 4(b)

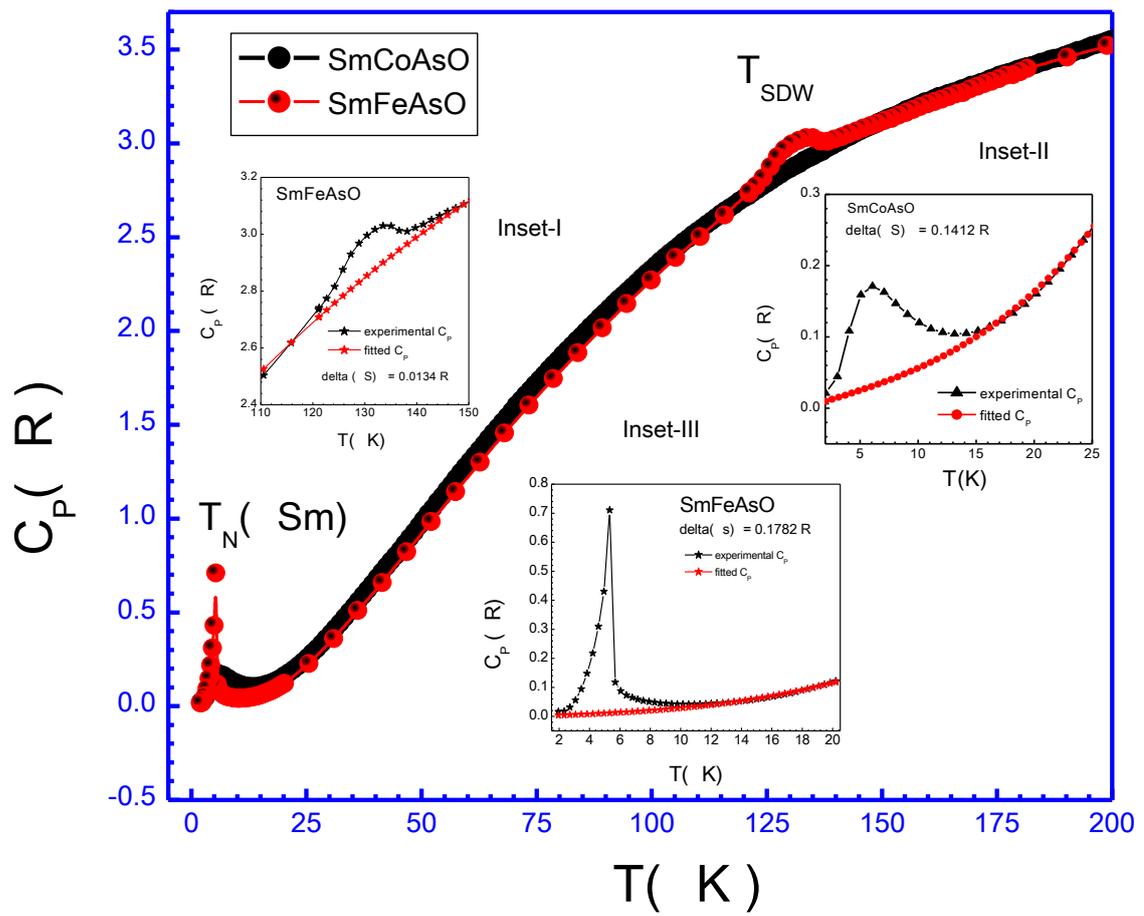

Fig. 5